**Title**

Real-time, noise and drift resilient formaldehyde sensing at room temperature with aerogel filaments

**Authors**

Zhuo Chen[1], Binghan Zhou[1], Mingfei Xiao[1], Tynee Bhowmick[1], Padmanathan Karthick Kannan[2], Luigi G. Occhipinti[1], Julian William Gardner[2], Tawfique Hasan[1]*

**Affiliations**

[1]Cambridge Graphene Centre, University of Cambridge, 9 JJ Thomson Ave., Cambridge CB3 0FA, UK

[2]School of Engineering, University of Warwick, Library Rd., Coventry CV4 7AL, UK

*Corresponding author. Email: th270@cam.ac.uk

**Abstract**

Formaldehyde, a known human carcinogen, is a common indoor air pollutant. However, its real-time and selective recognition from interfering gases remains challenging, especially for low-power sensors suffering from noise and baseline drift. We report a fully 3D-printed quantum dot/graphene-based aerogel sensor for highly sensitive and real-time recognition of formaldehyde at room temperature. By optimising the morphology and doping of printed structures, we achieve a record-high and stable response of 15.23 % for 1 part-per-million formaldehyde and an ultralow detection limit of 8.02 parts-per-billion consuming only ~130 µW power. Based on measured dynamic response snapshots, we also develop intelligent computational algorithms for robust and accurate detection in real time despite simulated substantial noise and baseline drift, hitherto unachievable for room-temperature sensors. Our framework in combining materials engineering, structural design and computational algorithm to capture dynamic response offers unprecedented real-time identification capabilities of formaldehyde and other volatile organic compounds at room temperature.

**Teaser**

Machine-intelligent aerogels enable highly sensitive, robust, and real-time recognition of formaldehyde at room temperature.

**MAIN TEXT**

**Introduction**

Formaldehyde is a known carcinogen as identified by the International Agency for Research on Cancer (*1*). It is a volatile organic compound (VOC) commonly found as an indoor air pollutant primarily emitted by wooden furniture and decorative materials (*2*). The adverse health implications of formaldehyde are significant even at low concentrations, encompassing symptoms such as coughing, fatigue, and can even potentially lead to fatal outcomes (*2*, *3*). These implications emphasise the urgent need to develop sensitive, instantaneous, and accurate gas sensors to detect formaldehyde, particularly in the presence of other interfering gases.

Metal oxide semiconductors (MOS) have been widely employed in detecting toxic, flammable, and irritating gases due to their diversity, low cost, simple structure, and stability (*4–6*). However, practical applications of MOS for accurate formaldehyde sensing are hindered owing to their high power consumption at elevated working temperatures, slow response and poor sensitivity at room temperatures (*7–9*). Crucially, they also have limited detection accuracy due to the cross-sensitivity towards other gases (*10–13*). To address this, substantial efforts have been dedicated to empowering gas sensors with artificial intelligence algorithms for gas classification. Nonetheless, these approaches are inefficient since they require measuring steady-state responses (*14–16*) or completing of particular testing sequences (*17–19*). Furthermore, cross-sensitivity for room temperature sensors remains unresolved. This is because the classification accuracy attained during training is unsustainable due to noise and baseline drift arising from large resistance and slow equilibration of defects (*20, 21*). These limitations have thus far rendered accurate measurement of formaldehyde unachievable for real-life applications.

Graphene-based aerogels are promising for room-temperature gas sensing due to their ultrahigh porosity and large surface area. However, their performance remains limited (*6, 22–24*) due to sub-optimised material and structural design. Indeed, the surface of chemically inert 3D porous graphene or reduced graphene oxide (rGO) requires appropriate sensitisation with other materials such as zero-dimensional MOS quantum dots (QDs) to absorb fast-diffusing gas molecules. Crucially, the rational design of the pores and skeleton structures of aerogels are critical for optimised sensor sensitivity, adsorption and desorption dynamics, and detection efficiency (*25, 26*). 3D printing allows engineering and fine-tuning the aerogel structures on multiple scales, providing additional dimensions for extracting dynamic and distinguishable features that are more resilient to noise and baseline drift, widely prevalent in gas sensors.

In this work, we report fully 3D-printed QDs/rGO-based aerogels for highly sensitive detection and real-time recognition of formaldehyde at room temperature. We quantitatively identify three critical factors that affect the gas sensing performance of the aerogels, namely, surface porosity, filament diameter, and surface doping of the printed structures. Through surface-reaction-coupled-diffusion modelling and empirical analysis, we show that printed aerogels with thinner filaments and larger surface porosity have better sensing response due to improved gas diffusion. The sensor with optimised aerogel structures eliminates the need for power-hungry components, such as heaters or light sources, and achieves an impressive response of 15.23 % for 1 parts-per-million (ppm) formaldehyde ($CH_2O$) in air at room temperature and an ultralow limit of detection (LOD) of 8.02 parts-per-billion (ppb) consuming only ~130 µW power. This is significantly below the workplace exposure limit (WEL) currently set in the UK at 2 ppm both for long-term (time-weighted average over eight hours) and short-term (averaged over ten minutes) exposure (*27*). We also develop gas recognition algorithms based on dynamic response features that do not require complete testing sequences or steady-state response and achieve extraordinarily accurate classification in real time. These features are inherently resilient to noise and baseline drift, enabling accurate classification of gas species in the presence of any Gaussian noise with standard deviation (SD) of up to 5 % or arbitrary baseline drift of up to ± 30 %. Our findings bring 3D-printed aerogel-based gas sensors a substantial step forward towards practical applications of machine-intelligence enhanced recognition.

## Results

### Ink formulation and 3D printing

$SnO_2$/GO hybrid materials are synthesized by a surfactant-assisted hydrothermal growth process as detailed in Fig. 1A and Methods. The $SnO_2$ QDs are uniformly distributed on GO sheets with a lateral size of ~1.4 μm (Fig. 1B, S2A), benefitting from the self-assembly process driven by the electrostatic interaction between the electropositive amino group of 6-aminohexanoic acid (AHA) and electronegative groups on GO sheets (*28*). The lattice fringes with an interplanar spacing of 2.66 and 3.32 Å (Fig. 1B) correspond to {101} and {110} planes of $SnO_2$, respectively. The average diameter of the QDs is 3.57 nm (Fig. S2B), less than twice the exciton Bohr radius of $SnO_2$. We employ a binary crosslinking agent strategy to formulate 3D-printable inks with suitable rheology. Specifically, copper ions ($Cu^{2+}$) coordinate with hydroxyl and carboxyl groups on GO sheets to increase viscosity (*29*) and simultaneously remove excess AHA ligands on the QDs surface through ligand exchange (*30*, *31*). Ascorbic acid (AA) is added to further increase the viscosity by partially reducing GO sheets and enhancing the π-π interaction between them (*32*, *33*). The formulated $SnO_2$/GO ink is stable and homogenous with GO sheets forming crosslinking networks (Fig. 1A).

The 0D/2D material-based aerogels are then fabricated by extrusion-based 3D printing as illustrated in Fig. 1C and Methods. The extrusion speed of the filament is controlled in accordance with the movement speed of the platform so that filaments can be laid onto the substrate smoothly with a diameter close to that of the extrusion nozzle. Diverse structures with tuneable morphologies can be fabricated and will be discussed later. Fig. 1D shows a typical aerogel gas sensor printed on a PCB substrate in a meander shape using an extrusion nozzle of 203 μm inner diameter (ID) that is subsequently freeze-dried and washed. The filaments are perpendicular to the bottom electrodes for better electrical contact and conductivity with a diameter of 250 μm and a separation distance of 500 μm (Fig. 1E). The measured filament diameter is comparable to the nozzle diameter because of the appropriately engineered storage modulus. The surface of the aerogel filament is porous with a porosity of 31 % calculated from a bespoke contrast-based image processing program; see Methods for details. The pores' area is marked in cyan colour on the right-hand side of Fig. 1F to demonstrate how the program identifies the pores and their boundaries.

The extrusion-based 3D printing method is versatile in fabricating various aerogel structures. For instance, aerogels with different layers can be easily fabricated by designing suitable printing patterns (Fig. 2A-C). Fig. 2D-F show the side view and Fig. 2G-I the top view of aerogel sensors with 1-layer, 2-layer, and 4-layer structures, respectively. When fabricating multiple layers, each additional meander-shaped layer is printed with a 90-degree rotation relative to the layer underneath. The filaments' shapes for the multilayer structure are well maintained due to the ink's high storage modulus close to $10^4$ Pa at low shear stress, as confirmed by the almost linear scaling of layer thickness from 220 μm to 420 μm and 820 μm when layer number increases from 1 to 2 and 4. By designing a suitable filament separation distance, the top layers can be well suspended without collapsing onto the bottom layers (Fig. 2K, L), thus providing good structural stability and unobstructed gas diffusion paths towards underlying layers.

We also fabricate aerogels with different surface porosities and filament sizes using extrusion nozzles with different inner diameters and extrusion speeds. While the filament diameter simply scales with the nozzle diameter, tuning surface porosity is more complicated. Based on our previous study (*25*), the surface porosity of printed aerogels is simultaneously controlled by the inks' crosslinking degree and the shear stress during printing. In principle, printing with a nozzle of smaller diameter at a higher flow rate will result in higher shear stress in the printing nozzle and smaller surface porosity in printed filaments. Specifically, high shear stress during high-speed extrusion in small nozzles causes flake alignment along the inner wall of the printing nozzle, leading to a non-porous surface, while crosslinking of the sheets hinders the process and preserves surface porosity. Fig. 2M shows that the shear stress applied by the nozzle wall to the aerogel surface is positively correlated with the extrusion speed. When the extrusion speed increases from 25 $\mu L \cdot min^{-1}$ to 100 $\mu L \cdot min^{-1}$ using the same nozzle (203 μm inner diameter), shear stress rises from 545 Pa to 600 Pa, resulting in smaller surface porosity of printed aerogels; see Fig. S3. Note that the aerogel fabricated with an extrusion speed of 100 $\mu L \cdot min^{-1}$ still preserves the porous surface despite the high shear stress during extrusion, owing to the counteracting crosslinking effects of $Cu^{2+}$ and AA. This corresponds to a printing speed of more than 3 $m \cdot min^{-1}$, enabling high speed and large-scale fabrication of aerogels with porous surface favourable for applications involving mass transfer.

Subsequently, we fabricate porous aerogels with different filament diameters using different extrusion nozzles. We carefully decrease the extrusion speed from 100 $\mu L \cdot min^{-1}$ to 25 $\mu L \cdot min^{-1}$ with nozzle diameter changing from 840 μm to 203 μm. This allows retaining similar surface porosities of various aerogels with different filament diameters (Fig. 2N). Additionally, we have chosen the centre-to-centre filament spacing to be at least twice the filament diameter to ensure good diffusion of gas molecules in between adjacent filaments. The gas sensing performance is thus not affected by the separation distance. Note that the 250-μm-thick film-like aerogel is specially fabricated by decreasing the distance between the extrusion nozzle (406 μm inner diameter) and the substrate such that adjacent filaments are flattened out and in touch with each other.

In addition to modulating the morphologies of aerogels, we also introduce metal doping of the QDs through a liquid-phase ligand exchange strategy. This process is achieved by immersing the freeze-dried aerogels in a metal salt solution in ethanol ($CuCl_2$, $NiCl_2$, $CoCl_2$) and then washing with hexane to remove any residual organic ligands, crosslinking agents and metal salt (Methods). This washing step enhances gas diffusion and expose the QDs' surface to gas molecules. These three metal ions are chosen because of their ability to crosslink GO sheets by coordination effects (*29*). Additionally, the oxide forms of these metals are widely employed in gas sensing applications (*34*). Ethanol is chosen as the solvent for ligand exchange due to its low surface tension and high polarity, which avoids detachment of the aerogels from the PCB substrate and initiates ligand exchange between metal ions and organic ligands on the QDs' surface. On the other hand, hexane is chosen as the solvent for washing due to its even lower surface tension and high vapour pressure at room temperature, eliminating the need for an extra freeze-drying step.

This also allows drying in the ambient air while preserving the aerogel surface porosity. Subsequent annealing partially removes oxygen functional groups of GO and oxidises the ligand-exchanged metal ions to metal oxide without affecting the overall aerogel structure. We note that washing with ethanol instead leads to a closed surface due to the collapse of pores' walls under surface tension during the drying process (Fig. S4). After the washing process, we confirm the successful removal of GO's intrinsic oxygen-containing functional groups, AHA and AA by the disappearance of relevant peaks in the infrared spectra of adsorption (Fig. S5). Raman spectra of resulted aerogels (Fig. S6) show a smaller area ratio of the D band to the G band ($A_D/A_G$) and a more prominent 2D peak, further confirming the removal of oxygen-containing functional groups and the partial recovery of the intrinsic graphene structure.

**Sensing performance**

We next assess the gas sensing performance of two-terminal aerogel gas sensors with different structures and chemical doping. The gas sensing experimental setup is detailed in Methods. The sensor response is defined as the fractional change in resistance $R$, $(R_a-R)/R_a \times 100\%$, where $R_a$ denotes the sensor's stabilised resistance in air. Fig. 3A shows the response curves of filament-structured aerogel gas sensors with the same filament diameter but different surface porosities towards 1 ppm formaldehyde. Upon gas exposure, the resistance $R$ increases slightly and then decreases, resulting in a stabilised positive response value. The initial resistance increases due to the interaction of $CH_2O$ molecules with the chemisorbed oxygen at lattice vacancies of the p-type surface dopant CuO. The reductant $CH_2O$ molecules donate electrons and consume holes in CuO, leading to the increase of conducting barriers and resistance (*35*). When more gas molecules diffuse into the aerogels, the interaction of $CH_2O$ molecules with the n-type $SnO_2$ QDs will dominate (*19*) Free electrons are donated by the chemisorbed $CH_2O$ molecules and transferred to the QDs uniformly distributed on rGO sheets. These electrons further transfer from QDs to rGO sheets that function as electron transport pathways towards the electrodes, contributing to increased current and decreased resistance. Note that for the aerogel sensor with 31 % surface porosity and 250 μm filaments, the response for 1 ppm $CH_2O$ reaches 15.23 % at the end of the gas exposure, among the highest value for MOX-based gas sensors ever reported at room temperature. The response value is positively correlated with the surface porosity, reducing to 10.6 % when the surface porosity decreases to 18 %. This can be attributed to the decrease in effective diffusion coefficient for filaments with lower surface porosity. Higher surface tension during printing promotes the alignment of flakes near the surface, leading to reduced surface porosity and increased tortuosity (*25*, *36*). Consequently, this combination results in a lower effective diffusion coefficient. Therefore, a higher surface porosity facilitates more efficient gas diffusion into the aerogels, leading to increased interactions between gas molecules and sensing materials.

On the other hand, we observe a negative correlation between the responses and filament diameters for filament-structured aerogel gas sensors. Figure 3B shows the response curves of filament-structured aerogels with similar surface porosity but different filament sizes under the exposure of 1 ppm $CH_2O$. The sensor constructed from filaments with a diameter of 250 μm has the highest response of 15.23 % at room temperature as discussed above. In contrast, sensors constructed from 400 μm and 600 μm filaments have response values of 12.63 % and 9.86 %, respectively. We also test a 250-μm-thick film-like aerogel sensor whose resistance changes by 10.94 %, inferior to filament-structured gas sensors with similar filament diameters. The champion device sustains discernible and repeatable response for lower concentration of $CH_2O$ down to 50 ppb; see Fig. S7A. From the low-concentration-response relationship in Fig. S7B, the LOD of $CH_2O$ can be determined to be as low as 8.02 ppb with the commonly assumed criterion of signal-to-noise ratio larger than 3. The noise-equivalent concentration is calculated by dividing the measured standard deviation of background noise level (0.005 %) by the slope of the regression line. The response towards 8.02 ppb formaldehyde, 3 times larger than the noise-equivalent concentration, is thus distinguishable and detectable by the readout circuit. The sensor shows good long-term stability within the testing period of 50 days with only 5.16 % baseline drift (Fig. S8A). This demonstrates the state-of-the-art gas sensing performance as shown in Table S3, without sacrificing long-term stability as suffered by sensors with noble metal doping. Furthermore, the power consumption during operation is merely ~130 μW under continuous operation in ambient air (190 kilohms, 5 V) or ~140 μW considering a full cycle of response and recovery in 1 ppm $CH_2O$, without any need of heaters or light sources. This is one order of magnitude smaller than the state-of-the-art gas sensors integrated with MEMS micro-heaters (*37*) and 50 % smaller than those integrated with micro light-emitting diodes (*38*), both requiring sophisticated fabrication processes. This demonstrates the superiority of 3D-printed filament-structured sensors that provide versatile control over structural morphology for improved gas diffusion and gas sensing performance.

To understand the structural dependence of gas sensing responses, we calculate the gas concentration profiles inside thin film by the surface-reaction-coupled gas diffusion model (*39*); see the Supplementary Text. We also expand this model to filament structures:

$$\begin{cases} C = C_\text{s} \dfrac{\cosh\left(\sqrt{\dfrac{k}{D}}x\right)}{\cosh\left(\sqrt{\dfrac{k}{D}}L\right)} & \text{for thin film} \\ \\ C = C_\text{s} \dfrac{I_0\left(\sqrt{\dfrac{k}{D}}r\right)}{I_0\left(\sqrt{\dfrac{k}{D}}R\right)} & \text{for filament structure} \end{cases} \quad (1)$$

where $C$ is the gas concentration inside the porous materials, $C_\text{s}$ is the concentration close to the surface, $D$ is the diffusion constant, $k$ is the surface reaction rate, $x$ is the distance from the bottom of the thin film with thickness $L$, and $r$ is the distance from the centre of the filament with radius $R$. Note that the pore sizes are much larger than that of the gas molecules such that molecular diffusion is the dominating mechanism instead of Knudsen diffusion. The calculated cross-sectional concentration profiles for a filament with 250 μm diameter and a layer of thin film with a thickness of 250 μm are compared and shown in Fig. 3E. The gas concentration remains at above 95 % of the surface concentration $C_\text{S}$ anywhere inside the filament, whereas the concentration inside thin film drastically decreases to 75 % at the bottom of the film (i.e. farthest from the surface). This demonstrates more efficient gas diffusion inside filament-structured gas sensors for which gas molecules can diffuse radially inwards from every direction, in sharp contrast with traditional thin-film gas sensors where gas molecules primarily diffuse from the top of the film only.

We then calculate and compare the concentration profiles for sensors with different filament diameters (Fig. 3D) or film thicknesses (Fig. 3F). Their diffusion constants and surface reaction rates are the same due to their similar pore sizes and operating temperature such that the gas concentrations close to surfaces are similar for these sensors. However, gas molecules cannot reach the deeper section of materials due to the chemical reaction at smaller depths. Thus, the normalised concentration profiles within the thicker structures are smaller for both filaments and films. The concentration profiles in the 250-μm-thick film-like aerogel are similar to the filament-structured aerogel with a 600 μm diameter, both smaller than that of the filament-structured sensor with a 250 μm diameter. The larger concentration profile in the filament-structured gas sensor with a smaller diameter ultimately leads to a superior relative resistance change and corresponds well with the trends in the measured sensor response values.

In addition to the structural dependence of response, we also find that doping the aerogels during the washing process leads to distinguishable gas sensing performance. Specifically, the sensor washed with $CuCl_2$ solution has a better response towards 1 ppm $CH_2O$ at room temperature, almost 3.4 times and 4.8 times higher compared to sensors washed with $NiCl_2$ and $CoCl_2$ respectively (Fig. 3C). This difference is likely due to different surface oxide materials (copper oxide, nickel oxide, cobalt oxide) and surface states formed on QDs' surface after the washing process (*30*), as confirmed by the X-ray photoelectron spectra analysis in Fig. S9. Particularly, CuO has been shown as a good catalyst for $CH_2O$ oxidation (*40*). The formed p-n junction at the $SnO_2$/CuO interface increases the sensor's resistance in air. Upon exposure to $CH_2O$, the depletion region responds sensitively to the released electrons from the reaction of $CH_2O$ molecules and oxygen species, resulting in a large modulation of resistance (*41*). Future optimisation could be carried out through theoretical calculation of adsorption energy and charge transfer capabilities of $CH_2O$ on different material interfaces. However, this is out of scope of our work here.

As ammonia ($NH_3$) and nitrogen dioxide ($NO_2$) are the other two major indoor air pollutants in addition to formaldehyde, we also explore the responses of aerogel sensors towards these two gases. Similar correlations of responses with surface porosity, structure, and chemical doping of aerogels are observed towards these two gas species (Fig. S10-12). Sensors with higher surface porosity, smaller filament diameter, and copper doping have better gas sensing performance. The dynamic response curves of the champion copper-doped filament-structured sensor with 31 % surface porosity and 250 μm filament diameter are shown in Fig. 3D. The response values at the end of target gas exposure are 15.23 % (1 ppm $CH_2O$), 14.25 % (1 ppm $NH_3$), 15.18 % (750 ppb $NO_2$), and 37.58 % (1 ppm $NO_2$), summarised in Fig. 3E. Though the sensor has very high sensitivity towards all three gas species, it becomes problematic in a mixed gas environment due to its poor selectivity. Indeed, it can be readily seen that the response values towards 1 ppm $CH_2O$, 1 ppm $NH_3$, and 750 ppb NO$_2$ are almost identical. One may combine aerogel sensors with different structural morphologies and doping to explore their differences in sensing response to classify various gas species. However, a precisely controlled gas flow procedure is required for measuring the response values at the end of gas exposure. Yet, in real life, such control of gas exposure is neither realistic nor cost-efficient. Moreover, one needs to wait for a certain amount of time before the sensing response stabilises, implying that if only using response as the metric, it is impractical to distinguish low-concentration highly reactive gases such as NO$_2$ from high-concentration less reactive gases such as CH$_2$O and NH$_3$. Thus, an appropriate and comprehensive data analysis technique that explores other distinguishable features from gas sensing results is needed for practical gas species recognition.

**Dynamic-feature-based recognition**

To address the above challenge, we develop a real-time gas species recognition algorithm based on a dynamic-feature-extraction approach. Different gas species have different reactivity and surface reaction rates, resulting in different sensitivity and sensing dynamics, such as the rate of resistance change during gas exposure or recovery. The latter can be used as features for classification through machine learning. Fig. 4A plots the response curves of the filament-structured sensor with 250 μm diameter during the gas exposure phase in solid lines, from which discernible differences in slope can be observed for different gas species. We then define the rate of change (RoC) as the ratio between the change of response $\Delta Rsp$ within a certain time interval or observation window $\Delta t$. The RoC calculated in multiple 30 s intervals of the response curves is shown in Fig. 4A as dashed lines. It is obvious that the RoC values for different gas species are almost always distinguishable, especially at the beginning of gas exposure. For instance, the RoC values are 0.026 %·s$^{-1}$, 0.016 %·s$^{-1}$, and 0.039 %·s$^{-1}$ for 1 ppm CH$_2$O, 1 ppm NH$_3$, and 750 ppb NO$_2$, respectively, during the 30 s observation window marked in the blue box. In addition to characterising the response dynamics with RoC, we also perform frequency domain analysis of response curves through discrete Fourier transform (DFT) to extract multi-dimensional features; see Fig. 4B. We denote the response values within the 30 s observation window as the time series $Rsp(t_0:t_{29})$, the complex DFT results as $X(f_0:f_{15})$, and their corresponding magnitudes as the frequency spectrum $M(f_0:f_{15})$. Since the sampling frequency of the sensor response is 1 Hz, the frequency spectrum covers from the DC value $M(f_0)$ to the magnitude at 0.5 Hz $M(f_{15})$. Fig. 4C shows the frequency spectra calculated from the response time series in the blue box in Fig. 4A. Similar to the RoC curves, distinguishable patterns can be observed from frequency spectrums for different gas species. For example, the pairs of the DC value $M(f_0)$ and the magnitude at fundamental frequency $M(f_1)$ are (2.68×10$^{-3}$, 2.56×10$^{-3}$), (4.44×10$^{-3}$, 1.63×10$^{-3}$), and (7.02×10$^{-3}$, 3.36×10$^{-3}$) for 1 ppm CH$_2$O, 1 ppm NH$_3$, and 750 ppb NO$_2$, respectively. We note that both the RoC and DFT spectra are dynamic features that only rely on a snapshot of data within a certain observation window and do not require the sensing response to reach a steady state. Very importantly, they dramatically reduce the time required for attaining discernible features for classification and enable real-time gas sensing and recognition.

To demonstrate the gas species recognition capability based on dynamic features, we construct aerogel sensor arrays of different structures and exploit the critical information embedded in their multidimensional response data. Specifically, four RoC values are calculated from the response data of four sensors, including three filament-structured aerogel sensors with filament diameters of 250, 400, and 600 μm and one 250-μm-thick film-like aerogel sensor. These four RoC values are regarded as features of one observation. Alternatively, eight characteristic DFT values are calculated by combining $M(f_0)$ and $M(f_1)$ of these four sensors' frequency spectra and regarded as features of one observation. The first observation is taken 60 s after the initial gas exposure to avoid instability and fluctuation of gas sensing responses at this stage. Subsequent observation windows with different $\Delta t$ can be randomly taken in real time to extract the corresponding dynamic features. It is worth noting that the gradual change of baseline resistance before gas exposure is much smaller than the extracted rate of change during gas exposure, thus will not affect the classification performance. More detailed discussion about baseline drift resilience is presented in the next section. The magnitudes at higher frequencies are not utilised here because of their smaller values and susceptibility to noise, but will be explored for counteracting the baseline drift in the next section. A dataset is next constructed with RoC values or characteristic DFT values calculated at random time intervals with a fixed length, with each interval denoted as one observation and labelled by gas species. The multi-dimensional dataset can be reduced to 2D space by the principal component analysis (PCA) algorithm that linearly transforms data to two principal component (PC) axes, PC1 and PC2, while preserving essential information such as data variation. A resultant 2D PCA map from the RoC dataset with $\Delta t$ of 30 s is plotted in Fig. S13, where clusters representing individual gas species are mostly separated, suggesting the possibility of identification of the tested analytes. It is also observed that these clusters are more separated during the initial exposure stage, and they move towards each other at the end of gas exposure when the response values stabilise and no longer have a discernible rate of change. This indicates that the RoC-based gas classification approach may be more suitable for fast and early-stage identification of just exposed gas species.

The PCs are used as features for training various classifiers, including linear discriminant, decision tree, k-nearest neighbours (kNN), Naïve Bayes, and support vector machine (SVM). The classification accuracies of these algorithms with five-fold validation are summarised in Table S1. The highest classification accuracy based on the RoC dataset is 97.5 % achieved by the SVM classifier with a $\Delta t$ of 30 s, which indicates the algorithm is able to predict the gas specie or class with 30 s of data at an accuracy of 97.5 %. The detailed classification performance encompassing both the accurate and confused predictions is shown in the confusion matrix plotted in Fig. 4D. On the other hand, most of the DFT-based classification algorithms achieve close to 100 % accuracy except for the linear discriminant. The confusion matrix and decision map using the SVM classifier on the DFT dataset are shown in Fig. 4E and 4F, respectively. The decision map plots out the data points representing different gas species or classes in different colours after PCA, namely $CH_2O$ (red), $NH_3$ (blue), and $NO_2$ (green). Clear boundaries are shown between each class indicating good separation and distinguishability from each other, which can be expressed as functions of 2D PC coordinates and readily used as the classification criteria for different gas species. We also evaluate the classification accuracies without performing dimensional reduction in the first place (Table S1). We found that PCA does not affect the champion SVM and kNN algorithms based on DFT or RoC with $\Delta t$ of 30 s, indicating that features are significantly correlated and those containing critical information about gas species are conserved during the dimensional reduction process. We also find that the data points representing different gases on the decision map based on DFT features are more clustered at the initial stage of gas exposure and then separated from each other further and further as the gas sensing response saturates, as opposed to the RoC-based gas recognition. This is due to including the fundamental frequency $M(f_0)$ in the feature set, which depends on the average response values within the observation window. At the initial stage of gas exposure, $M(f_0)$ values for different gases are all close to 0, and therefore may lead to indistinguishable features and inaccurate gas recognition. Indeed, by including the sensor response values at 1 to 4 minutes after the target gas exposure, more overlapping of data points can be observed in the corresponding PCA map (Fig. S14), which marginally reduces the overall classification accuracy to 97.5 %.

Dynamic-feature-based gas recognition is not limited to structurally multiplexed sensor arrays but is also effective for aerogel sensor arrays with different chemical dopants. In this case, three RoC values and six characteristic DFT values are calculated from three sensors doped with copper, nickel, and cobalt during the washing process. Following the same PCA-assisted machine learning approach, various classifiers are evaluated with accuracies summarised in Table S2. We note that PCA significantly lowers the classification accuracies of most RoC-based classification algorithms but leaves most DFT-based classifications intact. The highest accuracy for RoC-based classification is 85.4 % without PCA (Fig. 4G) or 74.2 % with PCA using the SVM classifier with $\Delta t$ of 30 s, while DFT-based classification predicts the best with the kNN classifier and achieves 99.2 % accuracy (Fig. 4H) even with PCA. Clear separation of clusters and boundaries between classes can also be observed in the decision map for the latter model (Fig. 4I). We find that DFT-based gas classification outperforms the RoC-based approach for aerogel sensor arrays with different structures or that with different dopants. More valuable features extracted from DFT can provide extra information favourable for classification. In summary, our dynamic-feature-based gas recognition algorithms are applicable for data acquired within a short observation window and do not rely on the completion of certain testing sequences or response at the steady state, thus enabling real-time gas classification with high accuracy. While the current measurements are conducted in dry air only, our gas recognition algorithms may also be extended to tackle the humidity interference and decouple its influence on sensitivity and selectivity by measuring the sensor response at various humidity and classifying them into separate classes (*18*, *42*). Similarly, our approach may be extended to a wider range of flow rates and gas concentrations by acquiring a larger dataset containing sensor responses at various flow rates and concentrations, and then applying classifier training followed by a predictive regression model (*18*).

**Noise- and baseline drift-resilient recognition**

The multidimensional information embedded in dynamic-feature-based gas recognition algorithms has excellent potential in solving the two persistent and practical problems that hinder normal operation of gas sensors and shorten their lifetime, namely noise and baseline drift. They will result in overlaps of data clusters representing different classes in the feature space and impair the classification (*43*, *44*). While the 0D-2D material-based aerogel gas sensors have an extremely low noise level of 0.005 % standard deviation (SD) and excellent signal-to-noise ratio in the order of $10^3$, there is still non-negligible noise during operation from various noise sources. The sensor may also be subject to slow baseline drift during prolonged operation at room temperature. Herein, we simulate electrical noise and baseline drift by adding Gaussian noise of different SD and baseline drift to the original gas sensing curves, as shown in Fig. 5A. We held the assumption that the simulated noise would not affect the intrinsic sensing response except only the signal-to-noise ratio. We also assume that the baseline drift is additive and slow enough such that the accumulated drift would not further develop within the timeframe of exposure to the target gas and measurements. The simulated baseline drift and Gaussian noise can thus be added to the original sensor output as follows and discussed in depth in the Supplementary Text:

$$\begin{cases} R_d(t) = R \times (1 + \alpha t) \\ Rsp_{d,n} = \dfrac{R_0 - R_d}{R_0} + s \times N(0,1) \end{cases} \quad (2)$$

The nominal response with baseline drift is calculated against the original resistance in the clean air without any baseline drift $R_0$ to reflect the real application scenario that lacks any baseline compensation measures. The baseline drift $\alpha t$ varies from -30 % to 30 %, whereas the noise SD $s$ varies from the intrinsic value of 0.005 % to 5 % (Fig. S15). These data embedded with noise and baseline drift subsequently go through a dynamic-feature extraction process to calculate the RoC values and characteristic DFT frequencies (Fig. 5B). Meanwhile, we also test different observation windows varying from 1 s to 60 s with the first observation taken at 60 s after exposure to the target gases. As many as four characteristic DFT magnitudes, $M(f_0)$, $M(f_1)$, $M(f_2)$, and $M(f_3)$, along with RoC values are extracted and normalised to zero mean and unit variance before being fed to various classifiers as discussed in the previous section (Fig. 5C). This step is to ensure that different features are kept at the same scale and weight during the PCA process. It is vital in maintaining good baseline-drift resilience since the absolute feature values will inevitably change as the baseline drifts (as discussed in the Supplementary Text).

The noise resilience performance of RoC-based algorithms is first evaluated with different observation windows, different noise magnitudes, and zero baseline drift. Every classifier is assessed for each combination of observation window and noise magnitude. The highest accuracy is summarised in Fig. 5D. We find that the observation window length plays a vital role in determining the classification accuracy, as briefly discussed previously. At a shorter window of 1 s or 5 s, the accuracy remains below 80 % for both sensor arrays with and without PCA. When the observation window increases to 10 s, the sensor array of different structures without applying PCA gives a satisfactory accuracy of 90.4 %. In contrast, the array of different dopants gives more than 90 % accuracy only when the observation window increases to 60 s, likely due to fewer extracted features than the array of different structures. The positive correlation between the observation window length and classification accuracy is attributed to the more substantial influence of signal noise and fluctuation on RoC-based features at shorter observation windows, within which the $\Delta Rsp$ is comparable to the noise level. This dependence, when there is only intrinsic noise, is projected and depicted on the left wall in Fig. 5D. The added simulated noise, on the other hand, dramatically affects the classification accuracy and results in false predictions. Introduction of simulated Gaussian noise with SD of 0.05 %, 10 times the SD of the intrinsic baseline noise, reduces accuracies of all RoC-based classification using either array at any observation window to below 80 %. The highest accuracy at this noise level is 78.7 %, achieved by the array of different structures with an observation window of 60 s. Increasing the noise SD to 0.2 % will further decrease the classification accuracy to 50 % rendering the classification invalid. We note that performing dimensional reduction to two principal components before classification tends to reduce the classification accuracy for all cases, indicating the loss of information embedded in multidimensional features.

On the other hand, the DFT-based classification is found to be much more resilient to noise; see Fig. 5E. All results are based on $M(f_0)$ and $M(f_1)$ characteristic DFT magnitudes. At the intrinsic noise level, both sensor arrays can distinguish different gas species with >96 % accuracies, even at a 1-s observation window when no PCA is performed, demonstrating the ultrafast classification capability. These results are projected and depicted on the left wall. As the noise level increases, the classification accuracy at 1 s observation window start to decrease, yet the best one remains above 90 % until the noise SD reaches 0.5 % and above 80 % until the noise SD reaches 2 %. We plot these results on the right wall. Notably, by only taking 2 samples within any 1 s observation window, the sensor array of different structures can achieve a classification accuracy of 92.1 % even when the noise SD increases to 0.2 %. We further note that choosing the appropriate observation window is crucial for maintaining high accuracies with higher noise levels. The shortest observation windows required for achieving classification at 90 % or higher accuracies increase with higher noise levels, corresponding to 5 s, 30 s, and 60 s for 0.5 %, 1 %, and 2 % noise SD, respectively. This is because longer observation windows effectively smooth out Gaussian noise and reduce its contribution to the calculated $M(f_0)$ and $M(f_1)$, ultimately increasing the signal-to-noise ratio in the extracted dynamic features. We also find that the sensor array of different structures performs better than that of different dopants due to more available features. PCA generally reduces classification accuracy due to lost information during dimensional reduction. To test the algorithms' overall resilience to any noise levels, we first train and select the 5 most accurate DFT-based classification algorithms at a 5 % SD noise level as shown in Fig. 5F. The most accurate classification has an accuracy of 85.2 % by using the structurally multiplexed sensor array with the observation window of 60 s and without PCA. We then apply these five algorithms to unseen data with smaller noise levels and tested their corresponding classification accuracies. We find that the best algorithm trained at 5 % SD noise level maintains high classification accuracy (>90 %) for all lower noise levels, demonstrating superior resilience to high noise levels during practical applications.

Finally, we investigate the baseline-drift-resilience of gas recognition algorithms based on dynamic features. We first note that RoC-based algorithms are intrinsically resilient to slow and additive baseline drift that does not alter the dynamics of the response curves, as discussed in the Supplementary Text. Therefore, given an arbitrary baseline drift, their classification accuracies at the intrinsic noise level are fully resilient to the drift and approximately the same as Fig. 5D. We then select the most accurate classification algorithm trained at 0.05 % noise SD, namely the RoC-based classification using the structurally multiplexed array with the observation window of 60 s, and apply it for classifying unseen data at other noise levels and with arbitrary baseline drifts as shown in Fig. 5G. We find that the classification accuracies remain almost the same across the whole drift range when the noise level is low, whereas, at higher noise level, the accuracy scales with the baseline drift while maintaining above 90 % at 0.02 % SD noise level regardless of the drift value. This also indicates that the noise-resilience of RoC-based algorithms is either suppressed or enhanced depending on whether the baseline drifts negatively or positively. For DFT-based algorithms, the extracted $M(f_0)$ is closely related to the DC component of the response curves and changes as the baseline drifts, whereas higher-order DFT magnitudes are more dependent on the shape and dynamics of the response curves and thus selected as the features for classification. The classification accuracies of DFT-based algorithms using $M(f_1)$, $M(f_1{:}f_2)$, or $M(f_1{:}f_3)$ are summarised in Fig. 5H. The dotted lines denote the accuracy boundary of 90 %. Using magnitudes of high-order frequencies as features achieves over 80 % accuracy for noise SD as high as 0.1 %, less resilient to simulated noises than when $M(f_0)$ is included as features. Classification algorithms using structurally multiplexed sensor arrays outperform those using doping multiplexed sensor arrays by showing a larger region with accuracies greater than 90 %, while there are no apparent differences among different numbers of DFT magnitudes used. We select the most accurate classification algorithms trained at 0.1 % noise SD, namely the DFT-based classification using the structurally multiplexed array and DFT magnitudes of high-order frequencies with the observation window of 60 s. We apply them for classifying unseen data at other noise levels and with arbitrary baseline drifts, as shown in Fig. 5I. We find that the classification accuracies remain almost the same across the whole drift range when the noise level is low, whereas, at higher noise levels, the accuracy scales with the baseline drift while maintaining above 90 % at 0.05 % SD noise level regardless of the drift value. Furthermore, we verified the stability of the extracted RoC values and high-order DFT magnitudes in the real aerogel sensor when tested repeatedly within a 50-day period (Fig. S8). These dynamic features show minimum variation despite 5.16 % drift of the sensor baseline and 19.8 % drift of nominal response, enabling 100 % successful recognition of $CH_2O$ within the testing period. In conclusion, both RoC-based and DFT-based algorithms using DFT magnitudes of high-order frequencies can withstand substantial arbitrary baseline drifts even in noisy environments, greatly extending their practical use for long-term operations.

**Discussion**

To summarise, we demonstrate successful 3D printing of $SnO_2$/rGO 0D-2D material-based aerogels into various structures for high-performance formaldehyde sensing at room temperature with real-time (<30 s) detection and selective classification of interfering species. We establish a close relationship between the gas sensing performance and the tuneable structural morphologies achieved through 3D printing. As supported by modelling and experimental results, the filament-structured aerogel sensors outperform film-like aerogel sensors due to more efficient gas diffusion, achieving an ultralow LOD of 8.02 ppb with only ~130 μW power consumption. By arranging aerogel sensors into arrays of different structural morphologies or chemical doping, we are able to accurately identify and distinguish different gas species in real time upon gas exposure using rate-of-change and Fourier-spectrum-based machine learning approaches. Our DFT-based strategy shows superior noise resilience and only requires a short observation window, making it an excellent promising candidate for practical gas sensing applications. Additionally, RoC and DFT features are resilient to arbitrary additive baseline drift, ensuring robust and accurate classification of formaldehyde from interfering gas species regardless of electrical noise and baseline drift. By strategically leveraging the synergy of optimisations in materials, microstructure and computational algorithms, our approach presents a machine-intelligent gas sensing strategy that may finally unlock low-power, reliable and real-time identification of hazardous gases including VOCs for indoor environments.

## Materials and Methods

### Ink formulation and characterisations

The $SnO_2$/GO hybrid materials are synthesised following the surfactant-assisted hydrothermal growth process. First, graphene oxide (GO) dispersion in water is prepared by dispersing non-exfoliated GO paste (Sigma-Aldrich) in DI water with a concentration of 25 mg mL$^{-1}$, followed by rigorous mechanical stirring for 3 hrs. $SnO_2$ precursor is separately prepared by dissolving 4 mmol of tin chloride pentahydrate ($SnCl_4 \cdot 5H_2O$, Sigma-Aldrich) and 4 mmol 6-aminohexanoic acid (AHA, Sigma-Aldrich) in 30 mL DI water and sonicating the solution for 5 min. This precursor is subsequently mixed with 10 mL GO dispersion by sonication for 30 min and transferred to a PTFE-lined digestion vessel (4744, Parr Instrument) for hydrothermal growth at 140 °C for 3 hrs. The vessel is cooled down to room temperature in cold water, and the product is centrifuged at 4000 rpm and washed with DI water 3 times before being re-dispersed in water with a concentration of 15 mg·mL$^{-1}$. Finally, 50 mM of copper chloride ($CuCl_2$, Sigma-Aldrich) and 28 mM of ascorbic acid (AA, Acros Organics) are added into the as-synthesized suspension for crosslinking and gelation. The mixture is heated at 60 °C for 1 h in the oven and used within one day as ink for 3D printing. High-resolution transmittance electron microscopy (HRTEM, Tecnai F20) is performed to characterise the as-synthesized inks.

### 3D printing and characterisations of aerogels

The $SnO_2$/rGO aerogel gas sensors are fabricated with a custom-built 3D printing setup. The hybrid ink is first loaded into a 5 mL syringe connected with a dispensing tip, then extruded at a constant flow rate and printed onto gold-plated interdigitated electrodes with finger width and spacing of 0.15 mm on printed circuit board (PCB, FR-4, TG170). The PCB is then frozen in liquid nitrogen for 10 min and transferred to a freeze dryer (LyoQuest, Telstar) for overnight freeze-drying. The resulting aerogels are soaked in a metal salt solution for 1 hr to introduce additional metal doping. 100 mM of copper chloride ($CuCl_2$, Sigma-Aldrich), nickel chloride ($NiCl_2$, Sigma-Aldrich), and cobalt chloride ($CoCl_2$, Sigma-Aldrich) solutions in ethanol are used respectively for different kinds of metal doping through liquid-phase ligand exchange. The samples are washed twice with hexane, dried in air and then annealed at 160 °C for 3 hrs to further reduce GO before being ready for gas sensing measurements. Morphology analysis of as-synthesized aerogels is performed with scanning electron microscopy (SEM, FEI Magellan 400), while chemical properties characterisation is performed with Fourier-transform infrared spectroscopy (FTIR, Agilent Cary 660), Raman spectroscopy (Horiba LabRAM Evolution, 523 nm excitation), and X-ray photoelectron spectroscopy (XPS, Thermo NEXSA G2, monochromated Al Kα X-ray source of 1486.7 eV).

## Gas sensing measurements

Gas sensing measurements are carried out in a Kenosistec gas characterisation system. Four mass flow controllers (MFC) are used, with one regulating the gas flow of dry air as the balance gas and the other three regulating 1 ppm formaldehyde ($CH_2O$), 1 ppm ammonia ($NH_3$), and 1 ppm nitrogen dioxide ($NO_2$) respectively as the target gas species. The target gas concentration is modulated by regulating the flow rate ratio of the target gas to balance gas with the total gas flow rate maintained at 500 sccm. Sensor resistance is recorded once every second at a fixed reading voltage, and the measurement chamber environment is maintained at 25 °C and 1 atm. A standardised conditioning process is carried out before each measurement by pumping down the chamber pressure to $10^{-2}$ torr, heating the sensor to 40 °C for 10 min, and then purging with dry air for 120 min.

## Data processing

Two types of feature extraction techniques are performed on sensor dynamic response curves. The first involves calculating the rate of response change within a specific time interval during the target gas exposure phase, whereas the other includes performing discrete Fourier transform (DFT) on data within such time interval to acquire frequency domain spectrum. The multidimensional features are reduced to a 2D map using the principal component analysis (PCA) method and then fed to various classification algorithms. The surface porosity of aerogels is calculated by binarising the SEM images based on contrast and calculating the ratio of pores' area to the surface area of filament, with code in the Supplementary Materials. SolidWorks is used for computational flow dynamics simulation and calculating the shear stress of nozzle walls based on the measured inks' rheology. MATLAB and Origin are used for data processing and visualisation.

**References**
1. IARC working group on the evaluation of carcinogenic risks to humans, *Chemical agents and related occupations* (International Agency for Research on Cancer, Lyon, vol.100, 2012).
2. T. Salthammer, S. Mentese, R. Marutzky, Formaldehyde in the indoor environment. *Chem. Rev.* **110**, 2536–2572 (2010).
3. R. Golden, Identifying an indoor air exposure limit for formaldehyde considering both irritation and cancer hazards. *Crit. Rev. Toxicol.* **41**, 672–721 (2011).
4. S. Y. Jeong, J. S. Kim, J. H. Lee, Rational design of semiconductor-based chemiresistors and their libraries for next-generation artificial olfaction. *Adv. Mater.* **32**, 2002075 (2020).
5. J. H. Lee, Linear gas sensing with dielectric excitation. *Nat. Electron.* **3**, 239–240 (2020).
6. J. Dai, O. Ogbeide, N. Macadam, Q. Sun, W. Yu, Y. Li, B. L. Su, T. Hasan, X. Huang, W. Huang, Printed gas sensors. *Chem. Soc. Rev.* **49**, 1756–1789 (2020).
7. X. Liu, S. Cheng, H. Liu, S. Hu, D. Zhang, H. Ning, A survey on gas sensing technology. *Sensors (Switzerland).* **12**, 9635–9665 (2012).
8. M. E. Franke, T. J. Koplin, U. Simon, Metal and metal oxide nanoparticles in chemiresistors: does the nanoscale matter? *Small.* **2**, 301–301 (2006).
9. N. Joshi, T. Hayasaka, Y. Liu, H. Liu, O. N. Oliveira, L. Lin, A review on chemiresistive room temperature gas sensors based on metal oxide nanostructures, graphene and 2D transition metal dichalcogenides. *Microchim. Acta*. **185**, 1–16 (2018).
10. Z. Song, W. Ye, Z. Chen, Z. Chen, M. Li, W. Tang, C. Wang, Z. Wan, S. Poddar, X. Wen, X. Pan, Y. Lin, Q. Zhou, Z. Fan, Wireless self-powered high-performance integrated nanostructured-gas-sensor network for future smart homes. *ACS Nano*. **15**, 7659–7667 (2021).
11. Z. Chen, Z. Chen, Z. Song, W. Ye, Z. Fan, Smart gas sensor arrays powered by artificial intelligence. *J. Semicond.* **40**, 111601 (2019).
12. Y. K. Jo, S. Y. Jeong, Y. K. Moon, Y. M. Jo, J. W. Yoon, J. H. Lee, Exclusive and ultrasensitive detection of formaldehyde at room temperature using a flexible and monolithic chemiresistive sensor. *Nat. Commun.* **12**, 1–9 (2021).
13. M. Khatib, H. Haick, Sensors for volatile organic compounds. *ACS Nano*. **16**, 16 (2021).
14. J. Chen, Z. Chen, F. Boussaid, D. Zhang, X. Pan, H. Zhao, A. Bermak, C. Y. Tsui, X. Wang, Z. Fan, Ultra-low-power smart electronic nose system based on three-dimensional tin oxide nanotube arrays. *ACS Nano*. **12**, 6079–6088 (2018).
15. A. T. Güntner, V. Koren, K. Chikkadi, M. Righettoni, S. E. Pratsinis, E-nose sensing of low-ppb formaldehyde in gas mixtures at high relative humidity for breath screening of lung cancer? *ACS Sensors*. **1**, 528–535 (2016).
16. H. M. Fahad, H. Shiraki, M. Amani, C. Zhang, V. S. Hebbar, W. Gao, H. Ota, M. Hettick, D. Kiriya, Y. Z. Chen, Y. L. Chueh, A. Javey, Room temperature multiplexed gas sensing using chemical-sensitive 3.5-nm-thin silicon transistors. *Sci. Adv.* **3**, 1–9 (2017).
17. T. C. Wu, A. De Luca, Q. Zhong, X. Zhu, O. Ogbeide, D. S. Um, G. Hu, T. Albrow-Owen, F. Udrea, T. Hasan, Inkjet-printed CMOS-integrated graphene–metal oxide sensors for breath analysis. *npj 2D Mater. Appl.* **3**, 1–10 (2019).
18. O. Ogbeide, G. Bae, W. Yu, E. Morrin, Y. Song, W. Song, Y. Li, B. L. Su, K. S. An, T. Hasan, Inkjet-printed rGO/binary metal oxide sensor for predictive gas sensing in a mixed environment. *Adv. Funct. Mater.* **32**, 2113348 (2022).
19. W. Tang, Z. Chen, Z. Song, C. Wang, Z. Wan, C. L. J. Chan, Z. Chen, W. Ye, Z. Fan, Microheater integrated nanotube array gas sensor for parts-per-trillion level gas detection and single sensor-based gas discrimination. *ACS Nano*. **16**, 10968–10978 (2022).
20. A. C. Romain, J. Nicolas, Long term stability of metal oxide-based gas sensors for e-nose environmental applications: An overview. *Sensors Actuators, B Chem.* **146**, 502–506 (2010).
21. G. Müller, G. Sberveglieri, Origin of baseline drift in metal oxide gas sensors: effects of bulk equilibration. *Chemosensors*. **10**, 171 (2022).
22. S. Dolai, S. K. Bhunia, R. Jelinek, Carbon-dot-aerogel sensor for aromatic volatile organic compounds. *Sensors Actuators, B Chem.* **241**, 607–613 (2017).
23. X. Liu, J. Cui, J. Sun, X. Zhang, 3D graphene aerogel-supported $SnO_2$ nanoparticles for efficient detection of NO2. *RSC Adv.* **4**, 22601–22605 (2014).
24. Z. Meng, R. M. Stolz, L. Mendecki, K. A. Mirica, Electrically-transduced chemical sensors based on two-dimensional nanomaterials. *Chem. Rev.* **119**, 478–598 (2019).
25. B. Zhou, Z. Chen, Q. Cheng, M. Xiao, G. Bae, D. Liang, T. Hasan, Controlling surface porosity of graphene-based printed aerogels. *npj 2D Mater. Appl.* **6**, 1–8 (2022).
26. J. Feng, B. L. Su, H. Xia, S. Zhao, C. Gao, L. Wang, O. Ogbeide, J. Feng, T. Hasan, Printed aerogels: chemistry, processing, and applications. *Chem. Soc. Rev.* **50**, 3842–3888 (2021).
27. Health and Safety Executive, *EH40/2005 Workplace exposure limits* (The Stationery

Office, Norwick, ed. 4, 2020).
28. Y. Luo, S. Fan, Y. Luo, N. Hao, S. Zhong, W. Liu, Assembly of $SnO_2$ quantum dots on RGO to form $SnO_2$/N doped RGO as a high-capacity anode material for lithium ion batteries. *CrystEngComm*. **17**, 1741–1744 (2015).
29. H. Bai, C. Li, X. Wang, G. Shi, On the gelation of graphene oxide. *J. Phys. Chem. C*. **115**, 5545–5551 (2011).
30. Z. Song, S. Xu, J. Liu, Z. Hu, N. Gao, J. Zhang, F. Yi, G. Zhang, S. Jiang, H. Liu, Enhanced catalytic activity of $SnO_2$ quantum dot films employing atomic ligand-exchange strategy for fast response $H_2S$ gas sensors. *Sensors Actuators, B Chem.* **271**, 147–156 (2018).
31. M. A. Boles, D. Ling, T. Hyeon, D. V. Talapin, The surface science of nanocrystals. *Nat. Mater.* **15**, 141–153 (2016).
32. K. K. H. De Silva, H. H. Huang, M. Yoshimura, Progress of reduction of graphene oxide by ascorbic acid. *Appl. Surf. Sci.* **447**, 338–346 (2018).
33. Z. Sui, X. Zhang, Y. Lei, Y. Luo, Easy and green synthesis of reduced graphite oxide-based hydrogels. *Carbon N. Y.* **49**, 4314–4321 (2011).
34. H. J. Kim, J. H. Lee, Highly sensitive and selective gas sensors using p-type oxide semiconductors: Overview. *Sensors Actuators, B Chem.* **192**, 607–627 (2014).
35. X. Peng, J. Liu, Y. Tan, R. Mo, Y. Zhang, A CuO thin film type sensor via inkjet printing technology with high reproducibility for ppb-level formaldehyde detection. *Sensors Actuators, B Chem.* **362**, 131775 (2022).
36. H. Chen, A. Pei, J. Wan, D. Lin, R. Vilá, H. Wang, D. Mackanic, H. G. Steinrück, W. Huang, Y. Li, A. Yang, J. Xie, Y. Wu, H. Wang, Y. Cui, Tortuosity effects in lithium-metal host anodes. *Joule*. **4**, 938–952 (2020).
37. K. W. Choi, J. S. Lee, M. H. Seo, M. S. Jo, J. Y. Yoo, G. S. Sim, J. B. Yoon, Batch-fabricated CO gas sensor in large-area (8-inch) with sub-10 mW power operation. *Sensors Actuators, B Chem.* **289**, 153–159 (2019).
38. I. Cho, Y. C. Sim, M. Cho, Y. H. Cho, I. Park, Monolithic micro light-emitting diode/metal oxide nanowire gas sensor with microwatt-level power consumption. *ACS Sensors*. **5**, 563–570 (2020).
39. G. Sakai, N. Matsunaga, K. Shimanoe, N. Yamazoe, Theory of gas-diffusion controlled sensitivity for thin film semiconductor gas sensor. *Sensors Actuators, B Chem.* **80**, 125–131 (2001).
40. S. Momeni, F. Sedaghati, $CuO/Cu_2O$ nanoparticles: A simple and green synthesis, characterization and their electrocatalytic performance toward formaldehyde oxidation. *Microchem. J.* **143**, 64–71 (2018).
41. L. Y. Zhu, K. Yuan, J. G. Yang, H. P. Ma, T. Wang, X. M. Ji, J. J. Feng, A. Devi, H. L. Lu, Fabrication of heterostructured p-CuO/n-$SnO_2$ core-shell nanowires for enhanced sensitive and selective formaldehyde detection. *Sensors Actuators, B Chem.* **290**, 233–241 (2019).
42. T. C. Wu, J. Dai, G. Hu, W. B. Yu, O. Ogbeide, A. De Luca, X. Huang, B. L. Su, Y. Li, F. Udrea, T. Hasan, Machine-intelligent inkjet-printed α-$Fe_2O_3$/rGO towards $NO_2$ quantification in ambient humidity. *Sensors Actuators, B Chem.* **321**, 128446 (2020).
43. S. Brahim-Belhouari, A. Bermak, P. C. H. Chan, Gas identification with microelectronic gas sensor in presence of drift using robust GMM. *ICASSP, IEEE Int. Conf. Acoust. Speech Signal Process. - Proc.* **5** (2004), doi:10.1109/icassp.2004.1327240.

44. A. U. Rehman, A. Bermak, Drift-insensitive features for learning artificial olfaction in e-nose system. *IEEE Sens. J.* **18**, 7173–7182 (2018).
45. E. N. Fuller, K. Ensley, J. C. Giddings, Diffusion of halogenated hydrocarbons in helium. The effect of structure on collision cross sections. *J. Phys. Chem.* **73**, 3679–3685 (1969).
46. B. E. Poling, J. M. Prausnitz, J. P. O'Connell, *Properties of gases and liquids* (McGraw-Hill Educ., New York, ed. 5, 2001).
47. G. Sakai, N. Matsunaga, K. Shimanoe, N. Yamazoe, Theory of gas-diffusion controlled sensitivity for thin film semiconductor gas sensor. *Sensors Actuators, B Chem.* **80**, 125–131 (2001).
48. S. Bin Jo, H. J. Kim, J. H. Ahn, B. W. Hwang, J. S. Huh, D. Ragupathy, S. C. Lee, J. C. Kim, Effects of thin-film thickness on sensing properties of $SnO_2$-based gas sensors for the detection of $H_2S$ gas at ppm levels. *J. Nanosci. Nanotechnol.* **20**, 7169–7174 (2020).
49. A. C. Romain, J. Nicolas, Long term stability of metal oxide-based gas sensors for e-nose environmental applications: An overview. *Sensors Actuators, B Chem.* **146**, 502–506 (2010).
50. G. Müller, G. Sberveglieri, Origin of baseline drift in metal oxide gas sensors: effects of bulk equilibration. *Chemosensors*. **10**, 171 (2022).
51. S. Di, M. Falasconi, Drift correction methods for gas chemical sensors in artificial olfaction systems: techniques and challenges. *Adv. Chem. Sensors* (2012).
52. F. A. Akgul, G. Akgul, N. Yildirim, H. E. Unalan, R. Turan, Influence of thermal annealing on microstructural, morphological, optical properties and surface electronic structure of copper oxide thin films. *Mater. Chem. Phys.* **147**, 987–995 (2014).
53. M. Martínez-Gil, D. Cabrera-German, M. I. Pintor-Monroy, J. A. García-Valenzuela, M. Cota-Leal, W. De la Cruz, M. A. Quevedo-Lopez, R. Pérez-Salas, M. Sotelo-Lerma, Effect of annealing temperature on the thermal transformation to cobalt oxide of thin films obtained via chemical solution deposition. *Mater. Sci. Semicond. Process.* **107**, 104825 (2020).
54. A. Liu, G. Liu, H. Zhu, B. Shin, E. Fortunato, R. Martins, F. Shan, Hole mobility modulation of solution-processed nickel oxide thin-film transistor based on high-k dielectric. *Appl. Phys. Lett.* **108**, 233506 (2016).
55. D. Zhang, C. Jiang, J. Wu, Layer-by-layer assembled $In_2O_3$ nanocubes/flower-like $MoS_2$ nanofilm for room temperature formaldehyde sensing. *Sensors Actuators, B Chem.* **273**, 176–184 (2018).
56. Z. Bo, M. Yuan, S. Mao, X. Chen, J. Yan, K. Cen, Decoration of vertical graphene with tin dioxide nanoparticles for highly sensitive room temperature formaldehyde sensing. *Sensors Actuators, B Chem.* **256**, 1011–1020 (2018).
57. L. Zhao, Y. Chen, X. Li, X. Li, S. Lin, T. Li, M. N. Rumyantseva, A. M. Gaskov, Room temperature formaldehyde sensing of hollow $SnO_2$/ZnO heterojunctions under UV-LED activation. *IEEE Sens. J.* **19**, 7207–7214 (2019).
58. N. Jafari, S. Zeinali, Highly rapid and sensitive formaldehyde detection at room temperature using a ZIF-8/MWCNT nanocomposite. *ACS Omega*. **5**, 4395–4402 (2020).
59. D. Zhang, Y. Cao, Z. Yang, J. Wu, Nanoheterostructure construction and DFT study of Ni-doped $In_2O_3$ nanocubes/$WS_2$ hexagon nanosheets for formaldehyde sensing at room temperature. *ACS Appl. Mater. Interfaces*. **12**, 11979–11989 (2020).
60. H. Hu, H. Liang, J. Fan, L. Guo, H. Li, N. F. De Rooij, A. Umar, H. Algarni, Y. Wang, G. Zhou, Assembling hollow cactus-like ZnO nanorods with dipole-modified graphene nanosheets for practical room-temperature formaldehyde sensing. *ACS Appl. Mater. Interfaces*. **14**, 13186–13195 (2022).

## Acknowledgments

We acknowledge XPS support and consultation from Dr. Shaoliang Guan at Maxwell Centre, Cavendish Laboratory, University of Cambridge.

**Funding:** This research was supported by the Engineering and Physical Sciences Research Council (grant numbers EP/W024284/1, EP/W023229/1, EP/T014601/1). Specially, the XPS data collection was supported by the Henry Royce Institute for advanced materials through the Equipment Access Scheme enabling access to the Royce XPS facility at Cambridge (Cambridge Royce Facilities grant EP/P024947/1 and Sir Henry Royce Institute - recurrent grant EP/R00661X/1). BZ would like to acknowledge the CSC-Cambridge scholarship for financial support.

**Author contributions:** The project was initially conceptualised by ZC, BZ, and TH. ZC designed and conducted the experiments and analysed the data. BZ performed SEM characterisations. MX performed FTIR and Raman spectrometry characterisations. BZ and ZC wrote the MATLAB code for calculating surface porosity. ZC prepared the manuscript and figures. TH oversaw the project. All authors analysed and discussed the results and contributed to the final manuscript.

**Competing interests:** The authors declare that they have no competing interests.

**Data and materials availability:** All relevant data are available from the corresponding author on request. The code for surface porosity calculation can be found in the Supplementary Materials.

## Supplementary Materials

**Supplementary Materials PDF file includes:**

Supplementary Text
Figs. S1 to S15
Tables S1 to S3
References (45 to 60)

**Other Supplementary Materials for this manuscript include the following:**

Surface porosity calculation code

**Figures and Tables**

**Fig. 1. Hybrid material ink formulation and 3D printed aerogels.** (**A**) Schematic of hydrothermal synthesis of the $SnO_2$/GO hybrid ink with a photo showing its high viscosity ready for 3D printing. (**B**) HRTEM image of $SnO_2$ QDs with the inset showing the uniform distribution of $SnO_2$ QDs on a graphene sheet. (**C**) Illustration of the extrusion-based 3D printing process for fabricating aerogels with different filament sizes and layer numbers. (**D**) Photo and (**E**) SEM image of an aerogel gas sensor printed on PCB substrate. (**F**) High-magnification SEM image of the surface of fabricated aerogel with the pores' area on the right half marked in cyan colour.

**Fig. 2. Aerogels with different layers and surface porosity.** (**A-C**) 3D illustrations, (**D-F**) side-view photos, (**G-I**) top-view photos of one layer, two layers, and four layers, respectively. Scale bar 1 mm. (**J-L**) SEM images of aerogels of one layer, two layers, and four layers, respectively. Scale bar 200 μm (**M**) Surface shear stress and surface porosity of aerogels printed with different flow rates using the 203 μm diameter nozzle. (**N**) Feature size and surface porosity of aerogels printed with nozzles of different inner diameters.

**Fig. 3. Room temperature gas sensing characteristics and diffusion profile of aerogels.** (**A**) Response curves of filament-structured aerogel sensors with different surface porosities towards 1 ppm $CH_2O$. (**B**) Response curves of filament-structured aerogel sensors with different filament diameters and one film-like aerogel sensor towards 1 ppm $CH_2O$. (**C**) Response curves of filament-structured aerogel sensors with different dopants towards 1 ppm $CH_2O$. (**D**) Calculated gas concentration profiles due to diffusion inside filament-structured aerogels with different filament diameters and (**F**) inside film-like aerogels with different thicknesses, displaying both concentration values and contour plots. (**E**) Enlarged contour plots of the concentration profiles of filament with 250 μm diameter and film with 250 μm thickness, also showing the analytical expression. (**G**) Response curves of the filament-structured aerogel sensor with 250 μm diameter towards different gas species, and (**H**) the corresponding response values at the end of target gas exposure.

**Fig. 4. Extraction of dynamic features and classification of different gas species.** (**A**) Response values and the corresponding extracted RoC values during the target exposure stage. (**B**) Illustration of the discrete Fourier transform process. (**C**) Extracted DFT magnitudes at different frequencies. Based on the structural array of sensors, confusion matrices of the SVM classifiers with PCA based on (**D**) RoC features and (**E**) DFT features, respectively, (**F**) the decision surface of the latter classifier with the arrows denoting the directions of gas exposure. Based on the doping array of sensors, confusion matrices of (**G**) the SVM classifier without PCA based on RoC features and (**H**) the kNN classifier with PCA based on DFT features, (**I**) the decision surface of the latter classifier with the arrows denoting the progression of time after exposure to the target gas species.

**Fig. 5. Noise resilience and baseline-drift resilience investigations.** (**A**) Simulation of Gaussian noises with different standard deviations and simulation of baseline drift with different drift values. (**B**) Extraction of dynamic features using observation windows of different lengths. (**C**) Machine learning with different classifiers and prediction on unseen datasets. Training accuracies and noise resilience of (**D**) RoC-based and (**E**) DFT-based

classification algorithms with different observation windows. (**F**) Prediction accuracies of five best-of-all algorithms trained at the noise SD of 5 % using unseen datasets at smaller noise levels. Prediction accuracies and baseline-drift-resilience of best-of-all (**G**) RoC-based algorithm and (**I**) DFT-based algorithm using unseen datasets, with 90 % accuracies marked with the dashed lines. (**H**) Training accuracies and noise resilience of DFT-based classification algorithms using magnitudes of high-order frequencies. (**D**) to (**F**) share the same colour bar on the right of (**F**), and (**G**) to (**I**) share the same colour bar on the right of (**I**).

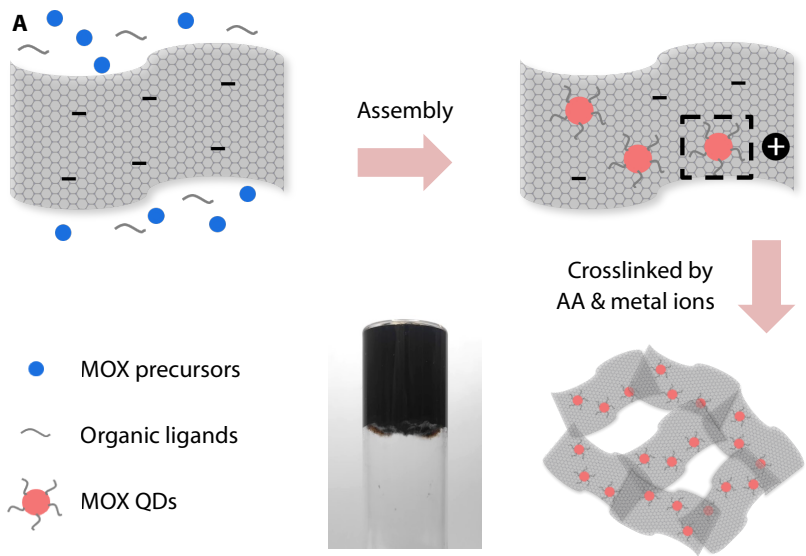
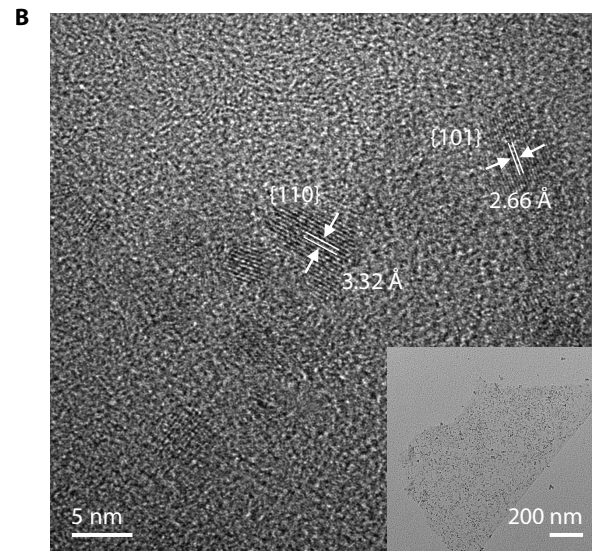
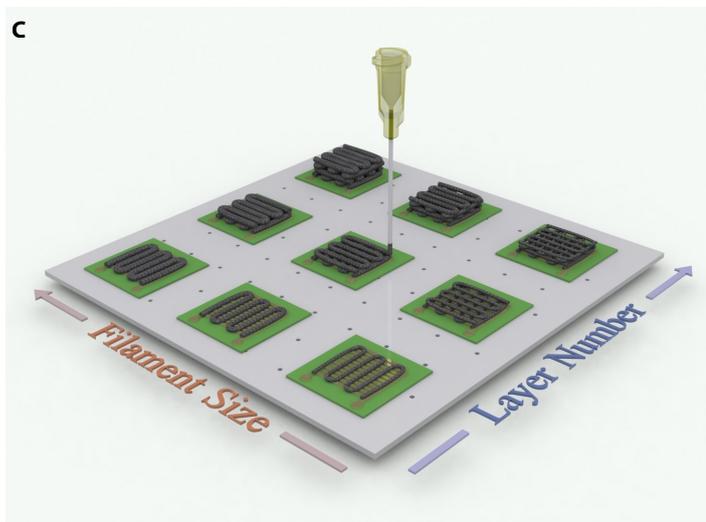
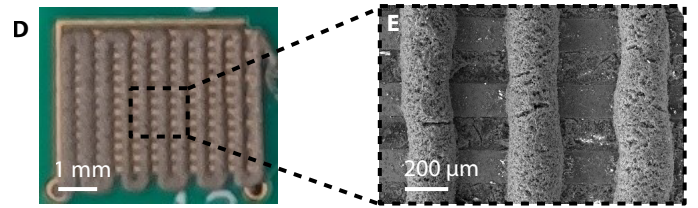
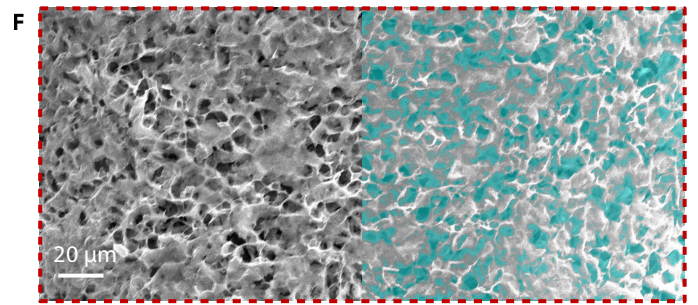

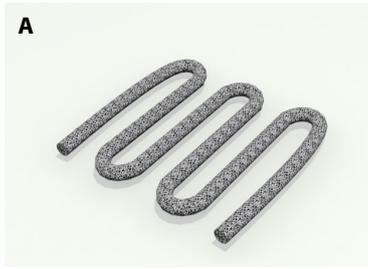 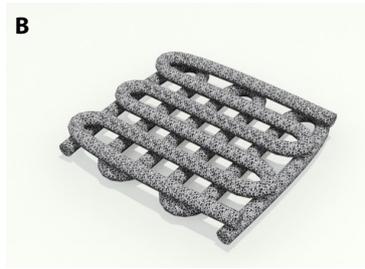 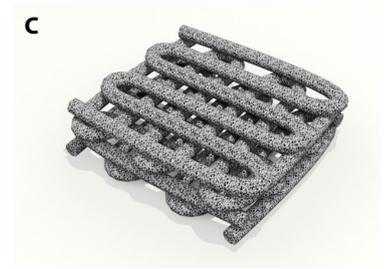
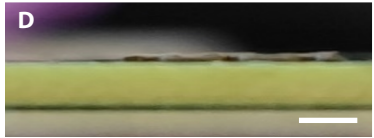 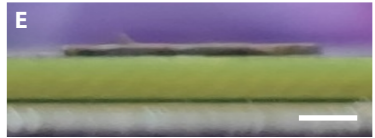 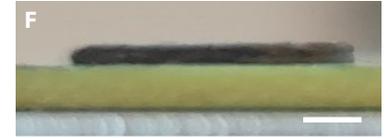
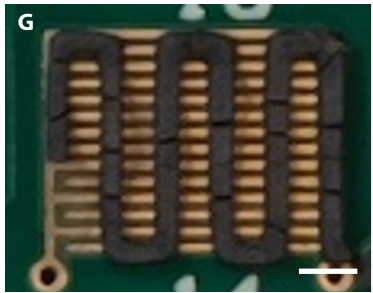 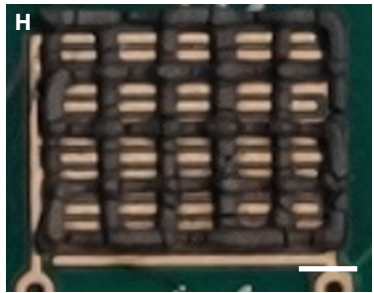 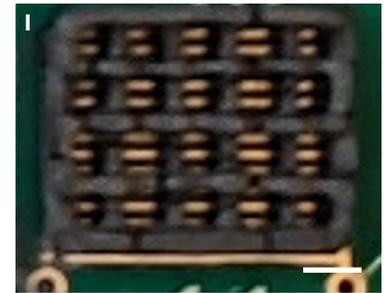
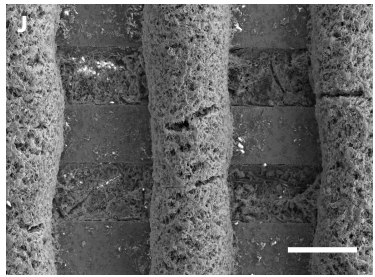 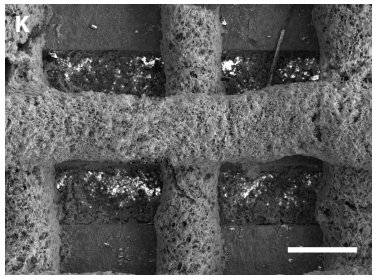 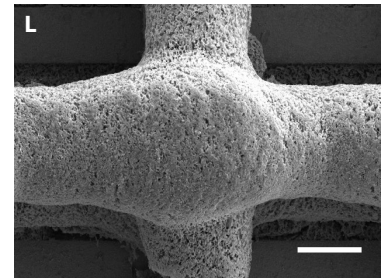
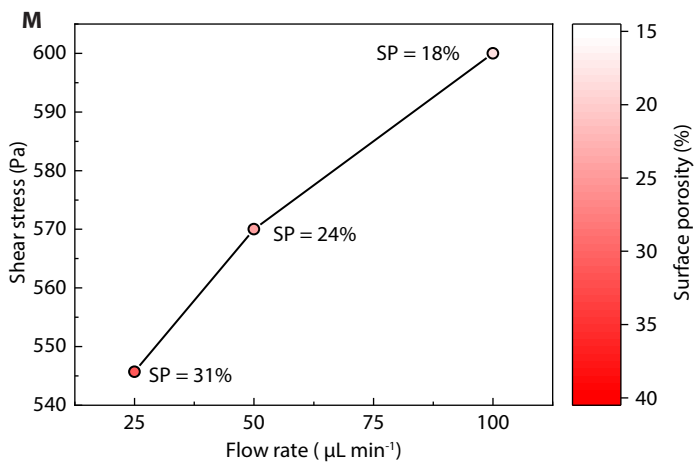 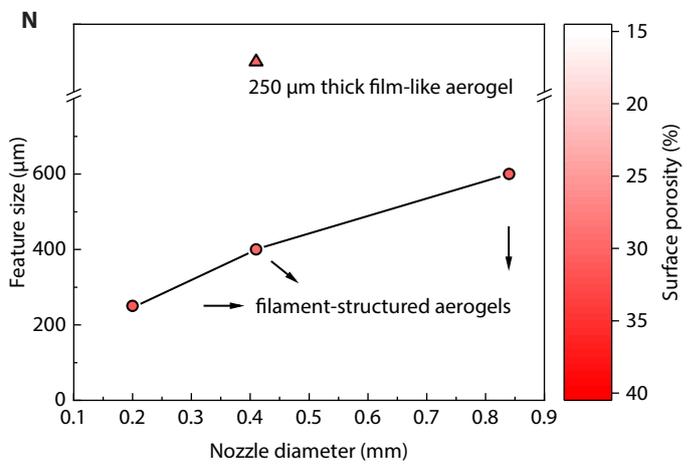

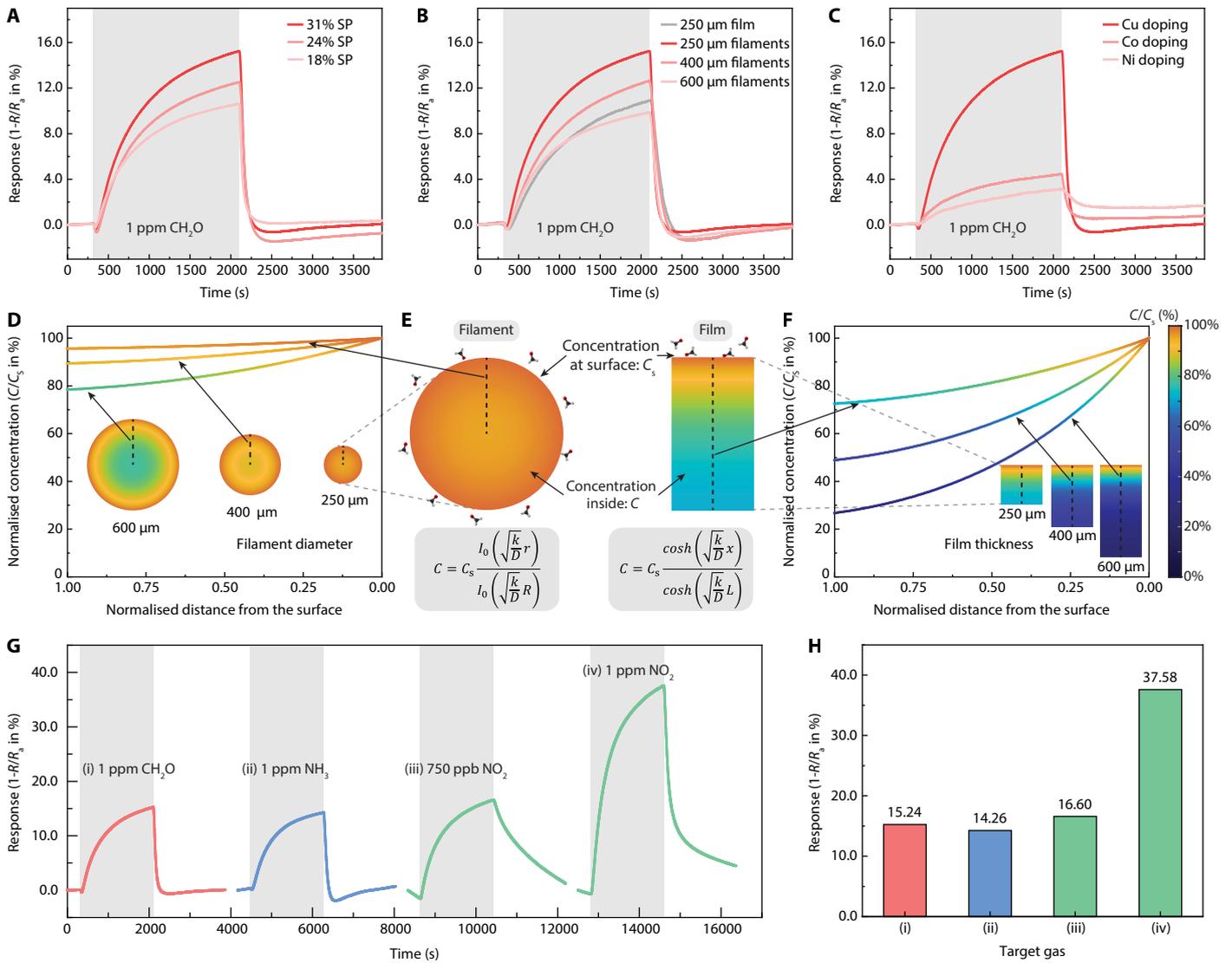

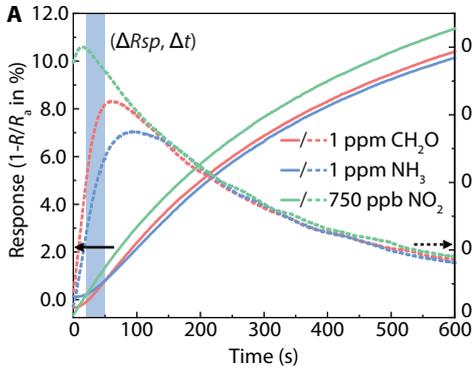
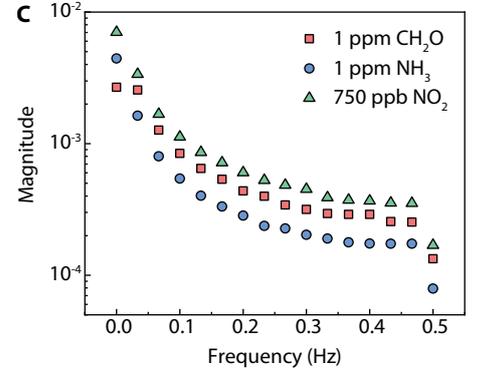
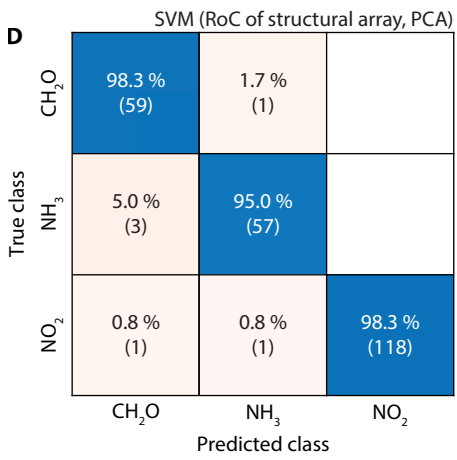
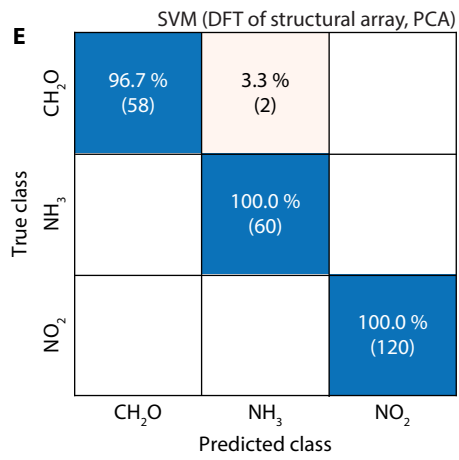
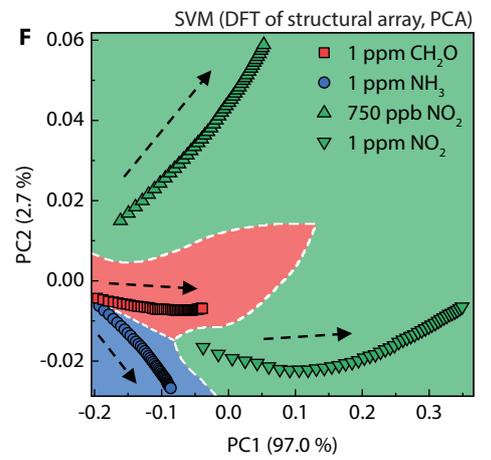
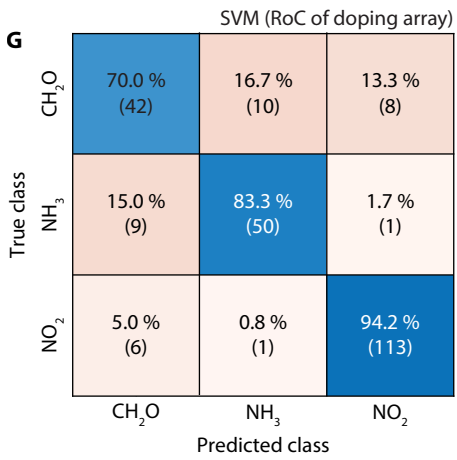
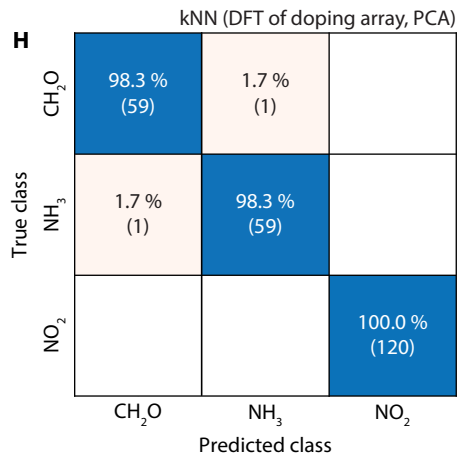
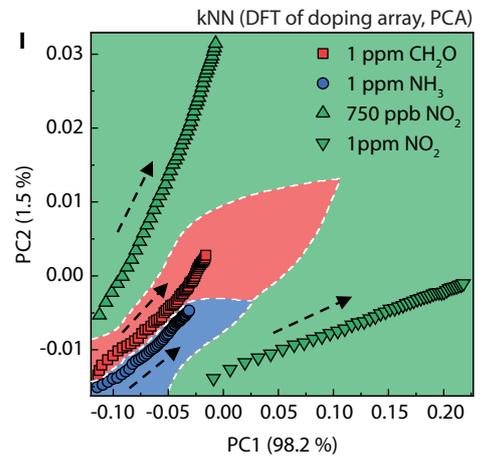

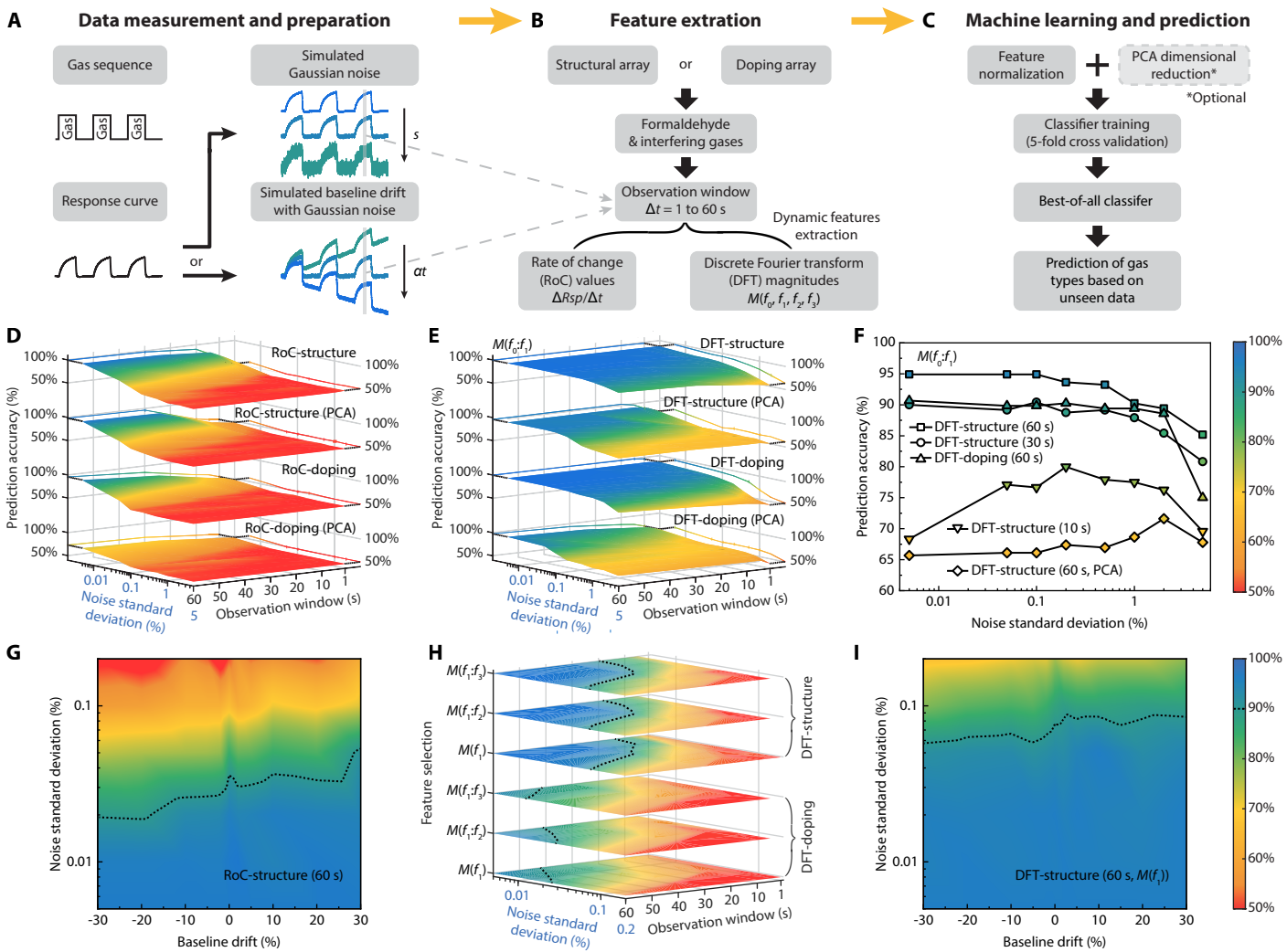